\documentclass[conference,10pt]{IEEEtran}

\usepackage{algorithm,algorithmic}
\usepackage{amsmath,amssymb,mathrsfs,amsbsy,amsthm}
\usepackage{cite}
\usepackage{graphicx,color}
\usepackage[dvipsnames]{xcolor}

\usepackage[hidelinks]{hyperref}
\usepackage{balance}
\usepackage{tikz,siunitx}
\usepackage{url}
\usepackage{subfigure}
\usepackage[inline]{enumitem}
\usepackage{booktabs}

\DeclareMathOperator*{\argmin}{arg\,min}

\definecolor{sblue}{RGB}{0,51,120}
\definecolor{sred}{RGB}{200,51,130}

\newcommand{\ex}{\mathbb{E}}
\newcommand{\tr}{\mathrm{tr}}

\newcommand{\fix}{{(F)}}
\newcommand{\val}{{(V)}}

\newcommand{\vbt}{{(V,\mathcal{B}_2)}}

\newcommand{\sign}{\mathrm{sign}}
\newcommand{\diag}{\mathrm{diag}}

\ifCLASSINFOpdf
\else
\fi

\begin{document}

\title{\huge Hybrid Constellation Modulation for Symbol-Level Precoding in RIS-Enhanced MU-MISO Systems}

\author{Yupeng Zheng, Yi Ma, and Rahim Tafazolli\\
	{6GIC, Institute for Communication Systems, University of Surrey, Guildford, UK, GU2 7XH}\\
		{Emails: (yz02263, y.ma, r.tafazolli)@surrey.ac.uk}\\
		}
\markboth{}%
{}

\maketitle

\begin{abstract}
The application of symbol-level precoding (SLP) in reconfigurable intelligent surfaces (RIS) enhanced multi-user multiple-input single-output (MU-MISO) systems faces two main challenges. First, the state-of-the-art joint reflecting and SLP optimization approach requires exhaustive enumeration of all possible transmit symbol combinations, resulting in scalability issues as the modulation order and number of users increase. Second, conventional quadrature amplitude modulation (QAM) exhibits strict constructive interference (CI) regions, limiting its effectiveness for CI exploitation in SLP. To address these challenges, this paper proposes a novel modulation scheme, termed hybrid-constellation modulation (HCM), which has a structure of superposed QAM and ASK sub-constellations (SCs). HCM extends the CI regions compared to QAM. Additionally, a two-stage reflecting and SLP optimization method is developed to support HCM. The proposed methods are designed for practical RIS with discrete phase shifts and has good scalability. Simulation results show that HCM achieves up to $1.5$ dB and $1$ dB SER gains over QAM with modulation order $16$ and $64$, respectively.\\
\end{abstract}


\section{Introduction}\label{Intro}
Reconfigurable intelligent surfaces (RIS) is a promising technology used to extend coverage of conventional wireless communication systems by adjusting the phase shifts of a large number of passive reflecting elements. Literature such as \cite{Wu2020,Zhou2022,Jiang2023} demonstrates that optimizing RIS to simultaneously serve multiple users in a multi-user multiple-input single-output (MU-MISO) system significantly increases spectral efficiency.

On the other hand, symbol-level precoding (SLP) is a technique that exploits both the channel state information (CSI) and the instantaneous transmitted symbols to design the transmit signal at the symbol level. Unlike conventional interference rejection precoding methods (e.g., zero-forcing (ZF) and regularized ZF (RZF)), which treat all interference as harmful, SLP exploits constructive interference (CI) among users, introducing additional degrees of freedom to optimize system performance in MU-MISO downlink\cite{Li2020}.

Recently, joint reflecting and SLP design in RIS-enhanced MU-MISO systems has been proposed in \cite{Liu2021}. The joint optimization algorithm shows significant gain in SER compared to linear precoding. However, the algorithm involves exhaustive enumeration of all possible transmit symbol combinations, making it computationally feasible only for low-rate modulation schemes (e.g., $4$-ary phase-shift keying ($4$-PSK) and $8$-PSK) with a small number of users. In addition, unlike practical RIS implementations which uses discrete phase shifts \cite{Wu2020}, \cite{Liu2021} assumes RIS elements with infinite phase-shift resolution. On the other hand, the strict constructive interference (CI) regions in conventional QAM constellations make it less likely for multi-user interference to fall within these regions, which is why QAM typically achieves lower gains in SLP compared to PSK \cite{cip_foundamental1,cip_foundamental3}.

Motivated by the above research gaps, this paper proposes a novel modulation technique, termed hybrid-constellation modulation (HCM), supported by a two-stage reflecting and SLP optimization method to enhance the scalability for higher modulation orders and more users. HCM combines QAM and amplitude-shift keying (ASK) in a superposed constellation. Based on the fact that an ASK symbol is fully determined by its real part, the imaginary part can be used for CI exploitation in SLP. This characteristic of HCM largely extends the CI regions compared to QAM. Simulation results of SER in RIS-enhanced MU-MISO downlink with SLP show that HCM has up to $1.5$ dB and $1$ dB gains to QAM with modulation order $16$ and $64$, respectively.

\textit{Notation:} Vectors and matrices are denoted by boldface lowercase letters (e.g., $\mathbf{a}$) and uppercase letters (e.g., $\mathbf{A}$). $(\cdot)^T$ and $(\cdot)^H$ denote transpose and Hermitian. $\mathfrak{R}(\cdot)$ and $\mathfrak{I}(\cdot)$ denote the real and imaginary parts of a complex number. $\mathrm{diag}(\cdot)$ denotes a diagonal matrix constructed from a vector. $\mathbb{E}(\cdot)$ denotes the expectation. $\mathcal{CN}(0, \sigma^2)$ denotes a circularly symmetric complex Gaussian distribution with zero mean and variance $\sigma^2$. $\tr(\cdot)$ denotes the trace of a square matrix. $\lvert\cdot\rvert$ and $\lVert \cdot \rVert$ denote the element-wise absolute value and Euclidean norm. $\mathbf{I}$ and $\mathbf{1}$ denote the identity matrix and the all-one vector. $\succeq$ and $\preceq$ denote element-wise inequality of vectors.

\section{System Model and Symbol-Level Precoding}
\subsection{System Model}
Consider a RIS-enhanced MU-MISO downlink system consisting of a BS, equipped with $M$ transmit antennas, serving $K$ single-antenna users. To enhance the quality of the communication link, a RIS composed of $N$ passive reflecting elements is deployed in the coverage area. Each element introduces a discrete phase shift to the incident signal, characterized by a finite set $\mathcal{F} = \left\{0, \frac{2\pi}{L}, \dots, \frac{2\pi(Q-1)}{Q}\right\}$, where $Q$ denotes the number of levels of discrete phase shifts.

The phase shift matrix is defined as
\begin{align}
	\boldsymbol{\Theta} = \text{diag}\{e^{j\theta_1}, e^{j\theta_2}, \dots, e^{j\theta_N}\},
\end{align}
where each $\theta_n \in \mathcal{F}$ denotes the angle of phase shift imposed by the $n$-th element. We assume that $\boldsymbol{\Theta}$ remains constant during the channel coherence time, which is a standard assumption in the literature (see, e.g., \cite{Liu2021,Wu2020}).

Let $\mathbf{h}_k \in \mathbb{C}^{M}$, $\mathbf{G} \in \mathbb{C}^{N \times M}$, and $\mathbf{f}_k \in \mathbb{C}^{N}$ denote the direct channel from the BS to user $k$, the channel from the BS to the RIS, and the reflected channel from the RIS to user $k$, respectively. We assume perfect CSI at the BS. The received signal at user $k$ can thus be expressed as
\begin{align}
	y_k = \sqrt{P_t/\xi}\left(\mathbf{h}_k^H + \mathbf{f}_k^H\boldsymbol{\Theta}\mathbf{G}\right)\mathbf{x} + v_k,
\end{align}
where $\mathbf{x}\in \mathbb{C}^{M\times 1}$ denotes the transmitted precoded signal vector, $\xi = \lVert \mathbf{x}\rVert^2$ is the rescaling factor which limits the total transmit power to $P_t$, and $v_k \sim \mathcal{CN}(0,\sigma^2)$ is additive white Gaussian noise (AWGN) at user $k$.

Assuming that all users employ the same modulation scheme and order, BS first modulates each user's random information into complex signals $s_k$. Then, through precoding, the symbol vector $\mathbf{s}=[s_1,\cdots,s_K]^T$ is multiplexed into $\mathbf{x}$ and simultaneously transmitted to all users.

The system model in matrix form for all users can be expressed as
\begin{align}
	\mathbf{y} &= \sqrt{P_t/\xi}(\mathbf{H} + \mathbf{F}\boldsymbol{\Theta}\mathbf{G})\mathbf{x} + \mathbf{v}\\
	& = \sqrt{P_t/\xi}\mathbf{H}_T\mathbf{x+v},\label{eq:sysmod2}
\end{align}
where $\mathbf{y} = [y_1,\dots,y_K]^T$, $\mathbf{H} = [\mathbf{h}_1, \dots, \mathbf{h}_K]^H$, $\mathbf{F} = [\mathbf{f}_1, \dots, \mathbf{f}_K]^H$, $\mathbf{v} = [v_1, \dots, v_K]^T$, and $ \mathbf{H}_T = \mathbf{H} + \mathbf{F}\boldsymbol{\Theta}\mathbf{G} $ denotes the total channel.

\subsection{Symbol-Level Precoding}

\begin{figure}[t]
	\centering
	\includegraphics[width=0.3\textwidth]{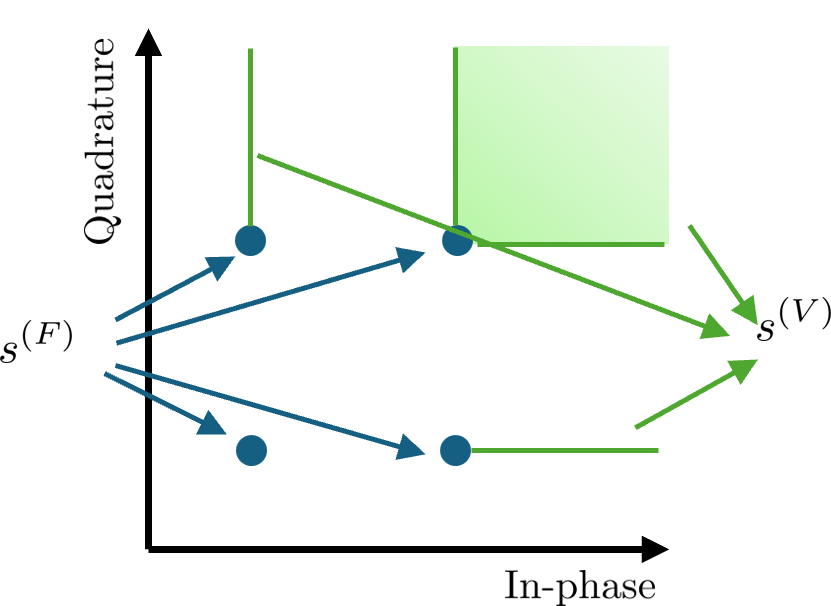}
	\caption{Fixed and variable parts of symbols in the first quadrant of $16$-QAM constellation. The fixed parts are denoted as blue dots, while the CI regions of the variable parts are denoted as green lines and area.}
	\label{fig:fixvalpart}
\end{figure}

SLP is a transmission technique designed to enhance the performance of MU-MISO systems by intentionally shifting the transmit symbols\cite{Li2020}. In SLP, a modulated symbol $s_k$ can be modeled as the sum of a fixed part $s^\fix_k$ and a variable part $s^\val_k$:
\begin{equation}
	s_k = s_k^\fix + s_k^\val.
\end{equation}
The SLP transmitter aims to accurately transmit the fixed part $s_k^\fix$ to the $k$-th user, while constraining $s_k^\val$ inside the feasible region, which is commonly known as the CI region\cite{Li2020}. Decomposing $s_k$ is convenient for formulating SLP optimization problems which will be shown in Section IV. 
Fig.~\ref{fig:fixvalpart} illustrates the fixed and variable part of the symbols in the first quadrant of $16$-QAM constellation.

\section{Hybrid-Constellation Modulation}
\begin{figure}[t]
	\centering
	\subfigure[$16$-HCM]{%
		\begin{minipage}[t]{0.49\columnwidth}
			\centering
			\includegraphics[width=\linewidth]{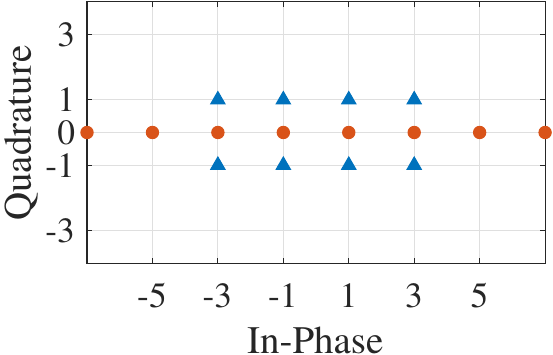}
			%			\vspace{0.5em}
		\end{minipage}%
	}
	\hfill
	\subfigure[$64$-HCM]{%
		\begin{minipage}[t]{0.49\columnwidth}
			\centering
			\includegraphics[width=\linewidth]{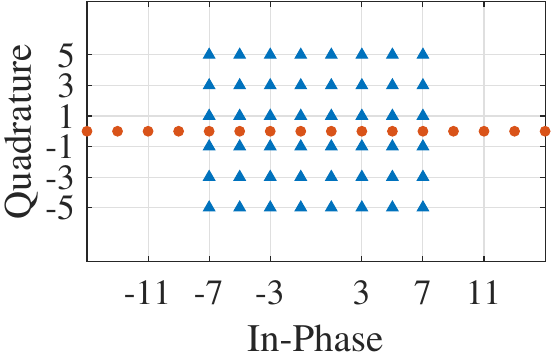}
			%			\vspace{0.5em}
		\end{minipage}%
	}
	\caption{\label{fig:txconst}Constellation diagrams of $16$-HCM and $64$-HCM, composed of QAM (blue triangles) and ASK (orange dots) SCs.}
	\vspace{-1em}
\end{figure}

We propose a modulation scheme named HCM to enhance SLP performance over QAM. HCM schemes are defined by a constellations with superposed ASK and QAM sub-constellations (SCs). Fig. \ref{fig:txconst} illustrates $16$-HCM constellation with $8$-ASK and $8$-QAM SCs and 64-HCM constellation with $16$-ASK and $48$-QAM SCs. The symbols in ASK and QAM SCs are marked as orange dots and blue triangles, respectively. The QAM SCs are in $I$-row-$J$-column rectangular structures. Since an ASK symbol is fully determined by its in-phase (real) component, the quadrature (imaginary) component can be exploited for optimization in SLP. 
%\begin{remark}
%	One may link HCM to the real-complex hybrid modulation (RCHM) proposed in \cite{rchm}. However, there are two main difference: First, an RCHM receiver needs side information to determined the expecting signal to be QAM or ASK, which HCM does not require. Second, both ASK and QAM symbols in HCM carry the same among of bits, which is not necessary in RCHM.
%\end{remark}

\begin{figure}[t]
	\centering
	\subfigure[$16$-HCM]{%
		\begin{minipage}[t]{0.49\columnwidth}
			\centering
			\includegraphics[width=\linewidth]{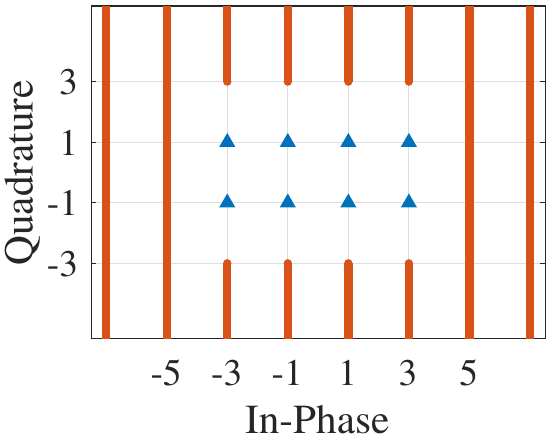}
			%			\vspace{0.5em}
		\end{minipage}%
	}
	\hfill
	\subfigure[$64$-HCM]{%
		\begin{minipage}[t]{0.49\columnwidth}
			\centering
			\includegraphics[width=\linewidth]{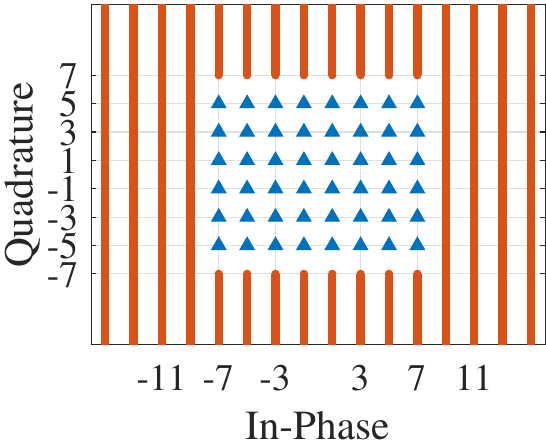}
			%			\vspace{0.5em}
		\end{minipage}%
	}
	\caption{\label{fig:rxconst}CI regions of $16$-HCM and $64$-HCM. Variability exists only in the quadrature components of ASK SCs, whose CI regions are marked as orange lines and half-lines.}
	\vspace{-1em}
\end{figure}

To facilitate the mathematical representation of the fixed and variable parts, we categorize the HCM symbols into 3 classes:
\begin{enumerate*}
	\item The symbols belong to the QAM SC are fixed in both real and imaginary parts and have variable parts equal to $0$.
	\item The symbols locate in the center of the ASK SC which overlap with the QAM SC have real fixed parts and constrained imaginary variable parts. The absolute values of the received imaginary parts need to be perturbed above $(I+1)$ to avoid collisions with the QAM SC. The CI regions of these symbols are illustrated as orange half-lines in Fig.~\ref{fig:rxconst}.
	\item The symbols from the ASK sub-constellation and lie outside of the QAM sub-constellation have real fixed parts and unconstrained imaginary variable parts. The CI regions of these symbols are illustrated as orange lines in Fig.~\ref{fig:rxconst}.
\end{enumerate*} 
Table \ref{tab:opcm_general} summarizes the fixed and variable parts of these 3 classes of HCM symbols, where $\mathcal{A}$ contain the values of the fixed parts and $\mathcal{B}$ represent the CI regions of the variable parts.

\section{Two-Stage Reflecting and SLP Optimization} 

In this section, we propose a two-stage optimization method for the discrete phase shift matrix $\boldsymbol{\Theta}$ and the SLP vector $\mathbf{x}$ with HCM. 
To avoid exhaustive enumeration over all possible symbol combinations such as the joint optimization algorithm in \cite{Liu2021} which has a computational complexity scaled exponentially with both the modulation order and $K$, we opt for a sequential strategy where $\boldsymbol{\Theta}$ is optimized first, followed by the optimization of $\mathbf{x}$ for the fixed $\boldsymbol{\Theta}$.
\subsection{Phase Shift Optimization}
It has been shown in \cite{cip_foundamental16} that SLP can be formulated as a symbol-perturbed zero-forcing (ZF) precoding problem, which indicates that the ZF precoder (i.e., the channel inverse) plays an essential role in SLP. Therefore, optimizing $\boldsymbol{\Theta}$ based on the ZF precoder can consequently enhance the performance of SLP. Assuming equal power allocation among all spatial streams, the problem of optimizing $\boldsymbol{\Theta}$ to minimize the power of the channel inverse can be formulated as
\begin{align}\label{eq:opt_theta1}
	(\mathrm{P1}):\argmin_{\boldsymbol{\Theta}}~&\tr\big((\mathbf{H}_T\mathbf{H}_T^H)^{-1}\big)\\
	\mathrm{s.t.}~&\theta_n \in \mathcal{F}, \forall n=1,\dots,N.
\end{align}
$\mathrm{P1}$ is a discrete optimization problem due to the finite set $\mathcal{F}$ of allowable RIS phase shifts. Solving $\mathrm{P1}$ via exhaustive search is computationally prohibitive, especially as $N$ grows large. To address this, we adopt the successive refinement method proposed in \cite{Wu2020}, which offers a low-complexity alternative and has been shown to achieve near-optimal performance. 

\begin{table}[t]
	\caption{Classification of HCM symbols.}
	\label{tab:opcm_general}
	\centering
	\renewcommand{\arraystretch}{0.4}
	%	\resizebox{0.45\textwidth}{!}{
		\begin{tabular}{cc}
			\toprule
			$s_{k}^{(F)}\in$ & $s_{k}^{(V)}\in$\\ 
			\midrule
			$\mathcal{A}_1$: QAM SC & $\mathcal{B}_{1}=\{0\}$ \\ 
			\addlinespace
			$\mathcal{A}_2$: ASK SC (central) & $\mathcal{B}_{2}=\{jz|z\in\mathbb{R},\lvert z\rvert\ge I+1\}$ \\ 
			\addlinespace
			$\mathcal{A}_3$: ASK SC (side) & $\mathcal{B}_{3}=\{jz|z\in\mathbb{R}\}$\\ 
			\bottomrule
		\end{tabular}
		%}
	\vspace{-1em}
\end{table}

\subsection{Optimization of SLP with HCM}
After $ \boldsymbol{\Theta} $ has been determined, BS can then optimize the transmit vector $ \mathbf{x} $ for the total channel $ \mathbf{H}_T $. To facilitate the discussion, we first introduce the widely-linear (WL) decomposition. The WL form of a complex vector $ \mathbf{a} $ and matrix $ \mathbf{A} $ are defined as
\begin{equation}
	\begin{aligned}
		\tilde{\mathbf{a}}=\begin{pmatrix}\mathfrak{R}(\mathbf{a})\\
			\mathfrak{I}(\mathbf{a})
		\end{pmatrix},
	\end{aligned}
\end{equation}
and 
\begin{equation}
	\begin{aligned}
		\tilde{\mathbf{A}}=\begin{pmatrix}\mathfrak{R}(\mathbf{A}) & -\mathfrak{I}(\mathbf{A})\\
			\mathfrak{I}(\mathbf{A}) & \mathfrak{R}(\mathbf{A})
		\end{pmatrix},
	\end{aligned}
\end{equation}
respectively. Since variable parts only exist in the imaginary parts of the HCM symbols, the WL modulated symbol can be sorted into
\begin{equation}\label{eq:swlsort}
	\mathbf{\Phi} \tilde{\mathbf{s}} = \begin{pmatrix}
		\mathbf{\Phi}^\fix\\
		\mathbf{\Phi}^\val
	\end{pmatrix}\tilde{\mathbf{s}}=
	\begin{pmatrix}
		\tilde{\mathbf{s}}^\fix\\
		\tilde{\mathbf{s}}^\val
	\end{pmatrix},
\end{equation}
where $ \mathbf{\Phi} $ is a permutation matrix and $ \tilde{\mathbf{s}}^\fix $ and $ \tilde{\mathbf{s}}^\val $ collect all the fixed and variable symbols in $ \tilde{\mathbf{s}} $, respectively. The optimization problem of SLP can be described as maximizing the signal-to-noise ratio (SNR) of the received fixed parts while constraining the received noise-free variable parts inside CI regions. According to \cite{zf_gen_inv}, the WL transmit vector with SLP can be modeled as
\begin{equation}\label{eq:xtilde}
	\tilde{\mathbf{x}} = \mathbf{W}\tilde{\mathbf{s}}^\fix = (\tilde{\mathbf{H}}_T^{\fix\dagger}+\mathbf{P}_\perp\mathbf{U})\tilde{\mathbf{s}}^\fix,
\end{equation} 
where $ \tilde{\mathbf{H}}_T^\fix = \mathbf{\Phi}^\fix \tilde{\mathbf{H}}_T $, $ \mathbf{A}^\dagger = \mathbf{A}^H(\mathbf{A}\mathbf{A}^H)^{-1}$ denotes the Moore-Penrose inverse, $\mathbf{P}_{\perp}=\mathbf{I}-\tilde{\mathbf{H}}_T^{(F)\dagger}\tilde{\mathbf{H}}_T^{(F)}$ is the orthogonal projection onto the null space of $\tilde{\mathbf{H}}_T^{(F)}$, and $\mathbf{U}$ is a matrix of real variables which is to be determined via optimization. According to \eqref{eq:sysmod2}, the rescaled WL receive vector can be written as
\begin{equation}\label{eq:sysmodWL}
	\sqrt{\xi/P_t}\tilde{\mathbf{y}} = \tilde{\mathbf{H}}_T\tilde{\mathbf{x}}+\sqrt{\xi/P_t}\tilde{\mathbf{v}}.
\end{equation}
Substituting \eqref{eq:xtilde} in \eqref{eq:sysmodWL} and extracting the fixed part gives
\begin{equation}\label{eq:yf}
	\sqrt{\xi/P_t}\tilde{\mathbf{y}}^\fix = \sqrt{\xi/P_t}\mathbf{\Phi}^\fix\tilde{\mathbf{y}} = \tilde{\mathbf{s}}^\fix + \sqrt{\xi/P_t}\tilde{\mathbf{v}}^\fix.
\end{equation}

The SNR of the $i$-th element in $\tilde{\mathbf{y}}^\fix $ can be calculated as 
\begin{equation}
	\label{eq:snri}
	\mathrm{SNR}_i = \frac{\xi\ex(\tilde{s}^{\fix 2}_i)}{P_t\ex(\tilde{v}_i^{\fix 2})} = \frac{2\xi\mathcal{E}_{\tilde{s}^\fix}}{P_t\sigma^2}
\end{equation}
where $\mathcal{E}_{\tilde{s}^\fix} = \ex(\tilde{s}^{\fix 2}_i)$ represents the average power of $\tilde{s}^{\fix}_i$, which is determined by the constellation. Notice that the only variable in \eqref{eq:snri} is $\xi$. Therefore, maximizing the $\mathrm{SNR}_i$ is equivalent to minimizing $ \xi $. On the other hand, since only the symbols in the second category in Table I need to be constrained in the imaginary part, we further extract the rows corresponding to these symbols from $ \tilde{\mathbf{H}}_T^\val = \mathbf{\Phi}^\val\tilde{\mathbf{H}}_T$ and denote it as $ \tilde{\mathbf{H}}_T^{\vbt} $. The noise-free receive signals which need to be constrained can be given by
\begin{equation}\label{eq:yv}
	\sqrt{\xi/P_t}\tilde{\mathbf{y}}^\vbt = \tilde{\mathbf{H}}_T^\vbt\mathbf{W}\tilde{\mathbf{s}}^\fix.
\end{equation}

Therefore, the SLP optimization problem with HCM can be formulated as
\begin{align}\label{eq:opt_theta2}
	(\mathrm{P2}):\argmin_{\mathbf{U}}~& \xi = \lVert(\tilde{\mathbf{H}}_T^{\fix\dagger}+\mathbf{P}_\perp\mathbf{U})\tilde{\mathbf{s}}^\fix\rVert^2\\
	\mathrm{s.t.}~\lvert \tilde{\mathbf{H}}_T^\vbt&(\tilde{\mathbf{H}}_T^{\fix\dagger}+\mathbf{P}_\perp\mathbf{U})\tilde{\mathbf{s}}^\fix\rvert \succeq (I+1)\mathbf{1},
\end{align}
Defining $\mathbf{U}\tilde{\mathbf{s}}^\fix = \mathbf{t}$, $\mathrm{P2}$ becomes
\begin{align}
	(\mathrm{P3}):\argmin_{\mathbf{t}}~& \lVert\tilde{\mathbf{H}}_T^{\fix\dagger}\tilde{\mathbf{s}}^\fix+\mathbf{P}_\perp\mathbf{t}\rVert^2\label{eq:p3obj}\\
	\mathrm{s.t.}~\lvert \tilde{\mathbf{H}}_T^\vbt&\tilde{\mathbf{H}}_T^{\fix\dagger}\tilde{\mathbf{s}}^\fix+\mathbf{P}_\perp\mathbf{t}\rvert \succeq (I+1)\mathbf{1}.
\end{align}
After the optimal $\mathbf{t}=\mathbf{t}^\star$ is obtained, any $\mathbf{U}$ which satisfies $\mathbf{U}\tilde{\mathbf{s}}^\fix = \mathbf{t}^\star$ is optimal. $\mathrm{P3}$ is a non-convex quadratic programme (QP) due to the element-wise absolute operation\cite{Boyd2004}. Nevertheless, it can be transformed into a series of convex sub-problems by adopting the concept of disjunctive programming\cite{Balas2018}. By introducing an auxiliary variable $\boldsymbol{\psi}$ and simplifying the expanded form of \eqref{eq:p3obj}, we obtain the following reformulation of $\mathrm{P3}$:
\begin{align}
	(\mathrm{P4}):&\arg\min_{\{\mathbf{t},\boldsymbol{\psi}\}}~\mathbf{t}^T \mathbf{P}_\perp^T \mathbf{P}_\perp \mathbf{t}\\
	\mathrm{s.t.}~&\boldsymbol{\Psi}(\tilde{\mathbf{H}}_T^\vbt\tilde{\mathbf{H}}_T^{\fix\dagger}\tilde{\mathbf{s}}^\fix+\mathbf{P}_\perp\mathbf{t})+(I+1)\mathbf{1} \preceq \mathbf{0},\label{eq:p4const}\\
	&\boldsymbol{\psi} \in \{-1,1\}^{N^{(V,\mathcal{B}_2)}},\quad\label{eq:p4const2}
\end{align}
where $\boldsymbol{\Psi} = \diag(\boldsymbol{\psi})$ and $N^{(V,\mathcal{B}_2)}$ denotes the dimension of $\boldsymbol{\psi}$. When $\boldsymbol{\psi}$ is fixed, $\mathrm{P4}$ becomes a linearly constrained QP (LCQP), which is convex and can be solved by toolboxes such as CVX \cite{Boyd2004}. Since the optimal $\boldsymbol{\psi}$ belongs to a finite set, the optimal solution can be obtained by traversing all possible values of $\boldsymbol{\psi}$. 

Since $\mathbf{t}$ has $2M$ entries and \eqref{eq:p4const} has $N^\vbt$ inequality constraints, the worst-case complexity of each LCQP in $\mathrm{P4}$ with interior-point method is of $\mathcal{O}\big((2M)^3+2M(N^\vbt)^2\big) $. Therefore, the total computational complexity of $\mathrm{P4}$ is of $\mathcal{O}\big(2^{N^\vbt}\big((2M)^3+2M(N^\vbt)^2\big)\big) $, which scales exponentially with the number of users. Next, we propose a sub-optimal heuristic algorithm which reduces the complexity by avoiding the exhaustive search of $\boldsymbol{\psi}$. 
\subsection{Sub-optimal SLP Algorithm for HCM}
Based on the intuition that shifting a symbol in the same direction as with the unconstrained optimizer (i.e., $\mathbf{t}=\tilde{\mathbf{H}}_T^{\fix\dagger}\tilde{\mathbf{s}}^\fix$) may incur lower power increase, we first estimate $\boldsymbol{\psi}$ as 
\begin{equation}\label{eq:signestimate}
	\hat{\boldsymbol{\psi}} = -\sign(\tilde{\mathbf{H}}_T^\vbt\tilde{\mathbf{H}}_T^{\fix\dagger}\tilde{\mathbf{s}}^\fix).
\end{equation} 
By substituting $\boldsymbol{\Psi}$ in \eqref{eq:p4const} with $\hat{\boldsymbol{\Psi}}=\diag(\hat{\boldsymbol{\psi}})$ and removing \eqref{eq:p4const2}, $\mathrm{P4}$ is simplified to the following LCQP
\begin{align}
	(\mathrm{P5}):\arg\min_{\mathbf{t}}~&\mathbf{t}^T \mathbf{P}_\perp^T \mathbf{P}_\perp \mathbf{t}\\
%	\mathrm{s.t.}~&\hat{\boldsymbol{\Psi}}\tilde{\mathbf{H}}_T^\vbt\mathbf{W}\tilde{\mathbf{s}}^\fix+(I+1)\mathbf{1} \preceq \mathbf{0}\label{eq:P5_2},\\
	\mathrm{s.t.}~\hat{\boldsymbol{\Psi}}\mathbf{P}_\perp\mathbf{t}+&\hat{\boldsymbol{\Psi}}\tilde{\mathbf{H}}_T^\vbt\tilde{\mathbf{H}}_T^{\fix\dagger}\tilde{\mathbf{s}}^\fix+(I+1)\mathbf{1}\preceq \mathbf{0},\label{eq:P5_2}
\end{align}
which has a significantly reduced total complexity of $\mathcal{O}\big((2M)^3+2M(N^\vbt)^2\big) $.

\section{Simulation Results}

\begin{figure}[t]
	\centering
	\includegraphics[width=0.35\textwidth]{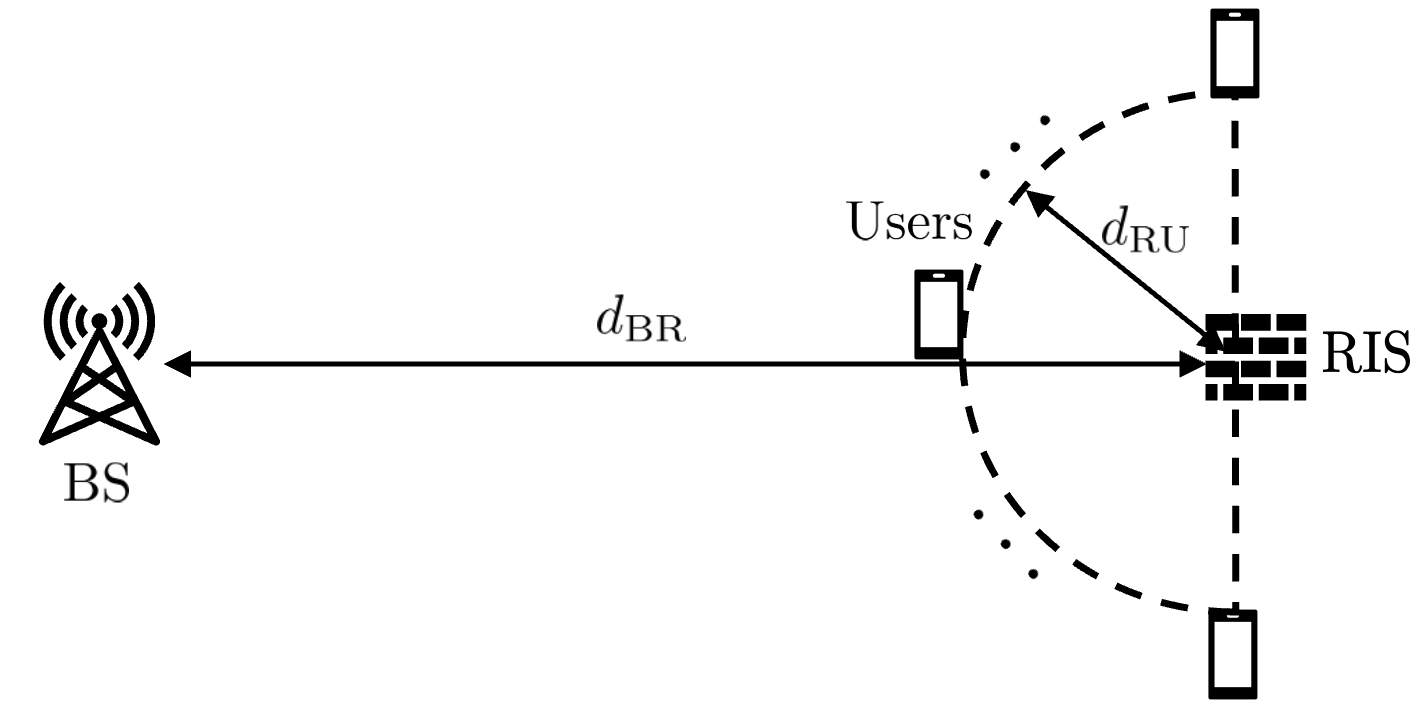}
	\caption{Simulation setup for RIS-enhanced MU-MISO. Users are uniformly distributed on a semicircle with a RIS located at the center.}
	\label{fig:irssetup}
\end{figure}

This section presents simulation results for SLP with the proposed HCM schemes in RIS-enhanced MU-MISO systems. The system setup is illustrated in Fig.~\ref{fig:irssetup}. The BS has $M=32$ uniform linear array (ULA) antennas and the RIS has $N=8\times 8 =64$ uniform rectangular array (URA) elements. Both the BS antenna spacing and the RIS element spacing are set to half the wavelength. The operating frequency is set to $3.5$ GHz. The distance between the BS and the RIS is $d_{\mathrm{BR}} = 100$ m, and the distance from the RIS to each user is $d_{\mathrm{RU}} = 10$ m. $K=32$ users are uniformly distributed on the half circle with the RIS placed at the center. Noise variance $\sigma^2 = -80$ dBm for all users. All channels ($ \mathbf{H,G,F} $) are modeled as independent and identically distributed (i.i.d.) Rician fading with a $\kappa$-factor equal to $3$ dB. The channel model is given by \cite{Liu2024} 
\begin{equation}
	\mathbf{A} = \sqrt{\frac{\kappa}{\kappa + 1}}\,\mathbf{A}^{\mathrm{LoS}} + \sqrt{\frac{1}{\kappa + 1}}\,\mathbf{A}^{\mathrm{NLoS}},
\end{equation}
where $\mathbf{A}^{\mathrm{LoS}}$ and $\mathbf{A}^{\mathrm{NLoS}}$ represent the line-of-sight (LoS) and non-line-of-sight (NLoS) components, respectively, with the former calculated from geometric positions and the latter generated from a standard complex Gaussian distribution. The path loss model can be given by $\mathrm{PL}(d) = C_0({d}/{d_0})^{-\alpha}$ \cite{Liu2021a}, where $C_0 = -30$ dB is the reference path loss corresponding to $d_0=1$ m. $\alpha$ denotes the path-loss exponent which is set to $3.5$, $2.5$, and $2.8$ for $\mathbf{H}$, $\mathbf{G}$, and $\mathbf{F}$, respectively. These configurations are based on existing works, e.g., \cite{Liu2021,Wu2020}.
\begin{figure}[t]
	\centering
	\subfigure[$16$-HCM]{%
		\begin{minipage}[t]{0.49\columnwidth}
			\centering
			\includegraphics[width=\linewidth]{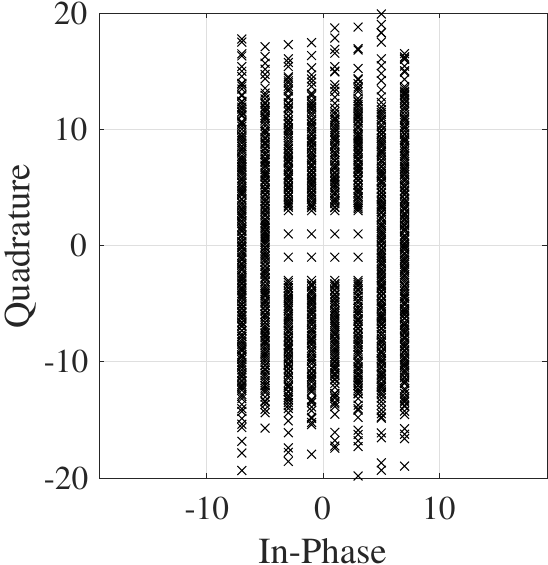}
			%			\vspace{0.5em}
		\end{minipage}%
	}
	\hfill
	\subfigure[$64$-HCM]{%
		\begin{minipage}[t]{0.49\columnwidth}
			\centering
			\includegraphics[width=\linewidth]{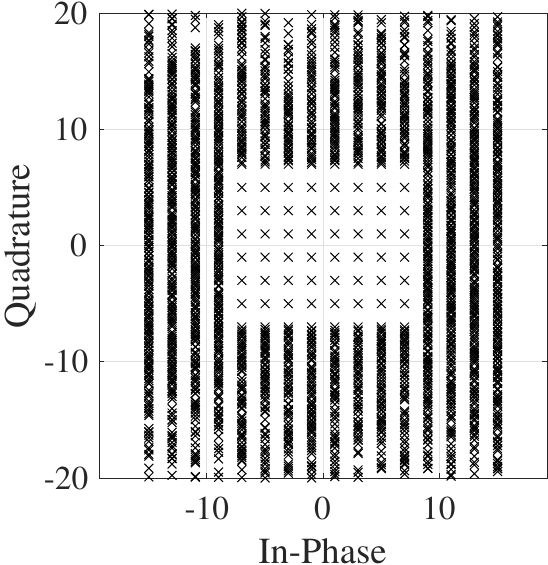}
			%			\vspace{0.5em}
		\end{minipage}%
	}
	\caption{\label{fig:scatte} Received noise-free $16$-HCM and $64$-HCM symbols. }
	\vspace{-1em}
\end{figure}
\subsection{Received Constellations of HCM}
Fig.~\ref{fig:scatte} presents the noise-free received constellation diagrams for the proposed $16$-ary and $64$-ary HCM schemes under SLP. The samples are collected via $1000$ Monte-Carlo trials. It can be observed that the ASK and QAM sub-constellations are well separated in the complex plane, allowing each user to detect the received symbol without requiring prior knowledge of whether the transmitted symbol belongs to ASK or QAM.
%\begin{figure}[t]
%	\centering
%	\includegraphics[width=0.49\textwidth]{Figures/plotScatter_spawc}
%	\caption{Constellation diagrams of noise-free received symbols of 16-HCM (left) and 64-HCM (right).}
%	\label{fig:scatte}
%\end{figure}

\begin{figure}[t]
	\centering
	\subfigure[Modulation order: $16$]{%
		\label{fig:ser_16}
		\begin{minipage}[t]{\columnwidth}
			\centering
			\includegraphics[width=0.9\linewidth]{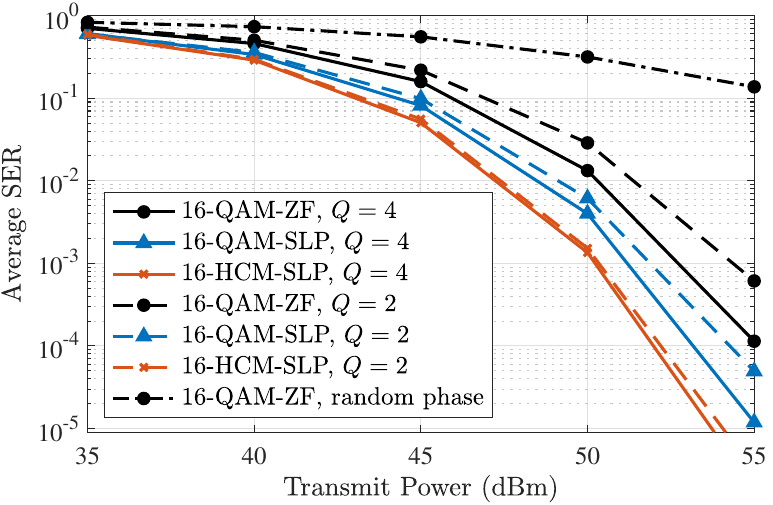}
			%			\vspace{0.5em}
		\end{minipage}%
	}
	\hfill
	\subfigure[Modulation order: $64$]{%
		\label{fig:ser_64}
		\begin{minipage}[t]{\columnwidth}
			\centering
			\includegraphics[width=0.9\linewidth]{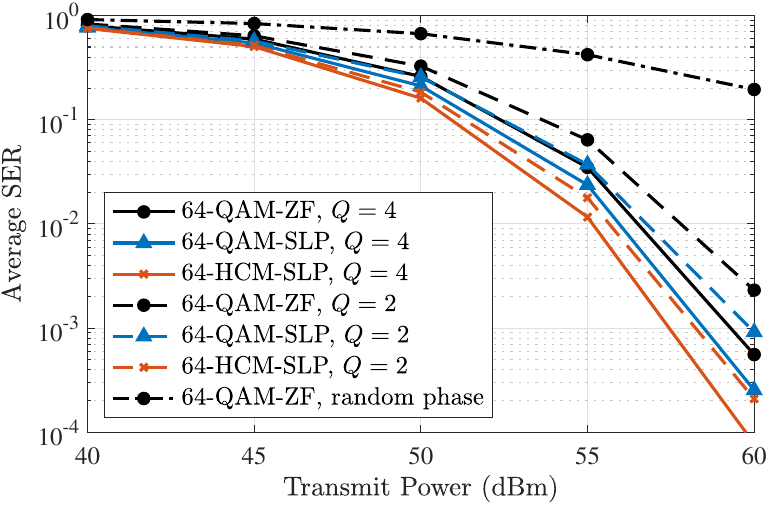}
			%			\vspace{0.5em}
		\end{minipage}%
	}
	\caption{\label{fig:ser}SER performance comparison in RIS-enhanced MU-MISO under different precoding schemes, modulations, and phase-shift levels with modulation order 16 and 64.}
	\vspace{-1em}
\end{figure}

\subsection{SER Performance}
In this subsection, we compare the SER performance of SLP with the proposed HCM scheme (labeled as HCM-SLP) to the following baselines: (i) ZF precoding with QAM (labeled as QAM-ZF) and (ii) SLP with QAM (labeled as QAM-SLP). The two-stage optimization method is applied for both HCM-SLP and QAM-SLP. The sub-optimal algorithm (\eqref{eq:signestimate}-\eqref{eq:P5_2}) is applied for HCM-SLP. For QAM-SLP, the closed-form algorithm in \cite{cip_foundamental1} is adopted. $Q$ is chosen as $2$ and $4$, which correspond to RIS phase-shift resolutions of 1-bit and 2-bit, respectively. Additionally, results with random phase shifts are also included for reference.

Fig. \ref{fig:ser} shows the simulation results of SER as functions of the transmit power in dBm for modulation order $16$ and $64$, respectively. QAM-ZF with successive phase refinement has over $8$ dB gain to QAM-ZF with random phase shift in both figures, which shows that the reflecting design already improves the system performance significantly before applying SLP. In Fig.~\ref{fig:ser_16}, $16$-HCM-SLP enhanced the gains of $16$-QAM-SLP from $3$ dB to $4.5$ dB for $Q=2$ and from $1.3$ dB to $2.5$ dB for $Q=4$ under $10^{-3}$ SER, thanks to the extended CI regions. Interestingly, $16$-HCM-SLP shows nearly identical SER performance for $Q=2$ and $Q=4$, which indicates that HCM-SLP is more robust to the low resolusion of RIS in this scenario. Similar performance gain is observed in Fig.~\ref{fig:ser_64}. $64$-HCM-SLP enhanced the gains of $64$-QAM-SLP from $1$ dB to $2$ dB for $Q=2$ and from $0.7$ dB to $1.5$ dB for $Q=4$ under $10^{-3}$ SER, respectively. $64$-HCM-SLP with $Q=2$ can reach the performance of $64$-QAM-SLP with $Q=4$, which shows that HCM can effectively compensate the degradation of RIS resolution.

%
%\begin{figure}[t]
%	\centering
%	\includegraphics[width=0.48\textwidth]{Figures/SER_RIS_32x64x32_L16}
%	\caption{SER performance comparison in IRS-aided MU-MISO with different precoding schemes, modulations, and phase-shift resolutions (modulation order: 16).}
%	\label{fig:ser_16}
%\end{figure}
%
%\begin{figure}[t]
%	\centering
%	\includegraphics[width=0.48\textwidth]{Figures/SER_RIS_32x64x32_L64}
%	\caption{SER performance comparison in IRS-aided MU-MISO with different precoding schemes, modulations, and phase-shift resolutions (modulation order: 64).}
%	\label{fig:ser_64}
%\end{figure}

\section{Conclusion}
In this paper, we proposed a novel modulation scheme named HCM and together with a two-stage optimization method to enable SLP in RIS-enhanced MU-MISO systems with higher modulation orders, a larger number of users, and discrete phase shifts. HCM has a unique constellation design which extends the CI regions compared to conventional QAM. The two-stage method avoids traversing all combinations of symbols which largely improves the scalability compared to the state-of-the-art algorithm. Simulation results show that HCM-SLP achieves up to $1.5$ dB and $1$ dB SER gains over QAM-SLP with modulation order $16$ and $64$, respectively. Additionally, HCM-SLP also achieves better performance than QAM-SLP even under lower RIS resolutions. 

\section{Acknowledgment}
This work was partially funded by the UKRI-EPSRC under Intelligent Spectrum Innovation (ICON) programme (APP55159) and has been filed in a patent with application number CN2025101692544.
%
%
%\begin{figure}[!ht]
%	\centering
%	\includegraphics[width=0.49\textwidth]{Figures/SER_RIS_32x64x32_L16}
%	\caption{.}
%	\label{}
%\end{figure}
%\begin{figure}[!ht]
%	\centering
%	\includegraphics[width=0.49\textwidth]{Figures/SER_RIS_32x64x32_L64}
%	\caption{.}
%	\label{}
%\end{figure}
\ifCLASSOPTIONcaptionsoff
\newpage
\fi

\bibliographystyle{IEEEtran}
\bibliography{./IEEEabrv, ./bibYZ.bib}
\end{document}